\newif\ifdraft
\newif\ifcomments

\ifdraft
    \documentclass[conference]{IEEEtran}
\else
    \documentclass[conference]{IEEEtran}
\fi

\usepackage{cite}
\usepackage[detect-all,group-separator={,}]{siunitx}
\usepackage{balance}
\usepackage{booktabs}
\usepackage{comment}
\usepackage{algorithmic}
\usepackage[inline]{enumitem}
\usepackage{graphicx}
\usepackage{caption}
\usepackage{lipsum}
\usepackage{multicol}
\usepackage{multirow}
\usepackage{textcomp}
\usepackage{lipsum}
\usepackage[dvipsnames]{xcolor}
\usepackage{caption}
\usepackage{tcolorbox}
\usepackage{subcaption}
\usepackage{hyperref}
\usepackage{float}
\usepackage{siunitx}
\usepackage{graphicx}
\usepackage{multirow}
\usepackage{booktabs}
\usepackage{tabularx}
\usepackage{fvextra}
\usepackage{csquotes}

\def\BibTeX{{\rm B\kern-.05em{\sc i\kern-.025em b}\kern-.08em
    T\kern-.1667em\lower.7ex\hbox{E}\kern-.125emX}}

\AtBeginDocument{%
  \providecommand\BibTeX{{%
    Bib\TeX}}}

\def\equationautorefname~#1\null{Eq.~(#1)\null}

\definecolor{amber}{rgb}{1.0, 0.75, 0.0}

\ifcomments
  \newcommand{\ian}[1]{{\textcolor{orange}{\textbf{Ian:}~\enquote{#1}}}}
  \newcommand{\kyle}[1]{{\textcolor{Mahogany}{\textbf{Kyle:}~\enquote{#1}}}}
  \newcommand{\ryan}[1]{{\textcolor{Magenta}{\textbf{Ryan:}~\enquote{#1}}}}
  \newcommand{\haochen}[1]{{\textcolor{amber}{\textbf{Haochen:}~\enquote{#1}}}}
   
  \newcommand{\fm}[1]{{\textcolor{red}{\textbf{FIXME:}~\enquote{#1}}}}
\else
  \newcommand{\ian}[1]{}
  \newcommand{\kyle}[1]{}
  \newcommand{\ryan}[1]{}
  \newcommand{\haochen}[1]{}
  \newcommand{\fm}[1]{}
\fi

\tcbuselibrary{minted,breakable,xparse,skins}

\definecolor{bg}{gray}{0.95}
\DeclareTCBListing{mintedbox}{O{}m!O{}}{%
  breakable=true,
  listing engine=minted,
  listing only,
  minted language=#2,
  minted style=default,
  minted options={%
    linenos,
    gobble=0,
    breaklines=true,
    breakafter=,,
    fontsize=\small,
    numbersep=8pt,
    #1},
  boxsep=0pt,
  left skip=0pt,
  right skip=0pt,
  left=15pt,
  right=0pt,
  top=3pt,
  bottom=3pt,
  arc=5pt,
  leftrule=0pt,
  rightrule=0pt,
  bottomrule=2pt,
  toprule=2pt,
  colback=bg,
  colframe=orange!70,
  enhanced,
  overlay={%
    \begin{tcbclipinterior}
    \fill[orange!20!white] (frame.south west) rectangle ([xshift=10pt]frame.north west); % 20pt
    \end{tcbclipinterior}},
  #3}

\begin{document}

\title{Icicle: Scalable Metadata Indexing and \\ Real-Time Monitoring for HPC File Systems}

\author{
  \IEEEauthorblockN{
    Haochen Pan\textsuperscript{*\dag},
    Ryan Chard\textsuperscript{\dag},
    Song Young Oh\textsuperscript{*}, \\
    Maxime Gonthier\textsuperscript{*\dag}, 
    Val\'{e}rie Hayot-Sasson\textsuperscript{\ddag},
    Geoffrey Lentner\textsuperscript{\S}, \\
    Joe Bottigliero\textsuperscript{*}, 
    Rachana Ananthakrishnan\textsuperscript{*\dag}, \\
    Kyle Chard\textsuperscript{*\dag},
    Ian Foster\textsuperscript{\dag *}
  }
  \IEEEauthorblockA{
    \textsuperscript{*}University of Chicago, Chicago, IL, USA
  }
  \IEEEauthorblockA{
    \textsuperscript{\dag}Argonne National Laboratory, Lemont, IL, USA
  }
  \IEEEauthorblockA{
    \textsuperscript{\ddag}\'{E}cole de technologie sup\'{e}rieure, Montr\'{e}al, QC, Canada
  }
  \IEEEauthorblockA{
    \textsuperscript{\S}Purdue University, West Lafayette, IN, USA
  }
}

\maketitle

\begin{abstract}

Modern HPC file systems can contain billions of files and hundreds of petabytes of data, making even simple questions increasingly intractable to answer. Traditional file system utilities such as \texttt{find} and \texttt{du} fail to scale to these sizes. While external indexing tools like GUFI and Brindexer improve query performance, they remain batch-oriented and unsuitable for heterogeneous, rapidly evolving environments.
We present Icicle, a scalable framework for continuous file system metadata indexing and monitoring. Icicle maintains a unified, up-to-date, and queryable view of file system state while supporting both periodic snapshot-based ingestion for bulk metadata updates and event-based ingestion for real-time synchronization from production systems such as Lustre and IBM Storage Scale. Built on Apache Kafka and Apache Flink, Icicle provides high-throughput, fault-tolerant, and horizontally scalable ingestion of metadata events into two complementary search indexes, enabling both individual file discovery and aggregate summary statistics by user, group, and directory.
This architecture enables efficient support for both coarse-grained administrative queries and interactive analytics over billions of objects. Our experimental evaluation on production-scale HPC datasets demonstrates order-of-magnitude throughput improvements over existing monitoring and indexing approaches, with tunable options for balancing consistency, latency, and metadata freshness.

\end{abstract}


\section{Introduction}
\label{sec:intro}
High performance computing (HPC) file systems can store tens of billions of files and hundreds of petabytes of data, driven by large-scale scientific experiments, extreme-scale simulations, and AI-driven workflows. For example, Oak Ridge National Laboratory (ORNL) hosts a 700 PB Lustre file system, Argonne National Laboratory has two 100 PB Lustre file systems, and the National Energy Research Scientific Computing Center (NERSC) maintains a 650 PB High Performance Storage System (HPSS) tape archive~\cite{braam_lustre_2019,olcf_frontier_storage_specs,alcf_eagle_grand_storage,nersc_storage_page}.

Massive datasets produced routinely by scientific applications must be reliably stored and managed across tiered and heterogeneous storage systems, each with independent namespaces and quotas, and distinct lifecycle policies, interfaces, and representations.
While this hierarchical design improves performance and optimizes data placement, it also fragments metadata across multiple systems, making user and administrative tasks increasingly complex and time-consuming.

Unfortunately, traditional utility tools such as \texttt{find} and \texttt{du} fail to scale to modern HPC file systems. Already in 2010, ORNL reported that a single \texttt{find} or \texttt{du} traversal on their 10 PB Spider file system could take more than 48 hours to complete~\cite{miller_monitoring_2010}. 
As file systems continue to grow beyond petabyte scales, administrators have turned to external metadata indexing systems to improve observability and conduct data lifecycle management. Modern solutions such as Grand Unified File Indexer (GUFI)~\cite{manno_gufi_2022} and Brindexer~\cite{paul_efficient_2020} construct external, database-backed indexes of file system metadata, enabling faster query performance than in-place scans. 
However, these systems remain fundamentally batch-oriented: they rely on periodic scans to update their databases and require parallel scripts to perform queries and analysis. 
As a result, obtaining even basic statistics such as per-user summaries or top-$k$ usage reports remains slow, resource-intensive, and operationally cumbersome.

Modern HPC file systems such as Lustre and IBM Storage Scale (GPFS)~\cite{schmuck_gpfs_2002} provide real-time metadata event feeds, yet few indexing systems exploit these capabilities to enable both real-time event monitoring and cost-aware, periodic indexing. At the same time, metadata consumers have widely varying freshness requirements: anomaly detection may need near-real-time updates, while usage trend analysis may tolerate  daily snapshots. Existing approaches, while effective for file counting or hierarchical aggregation, are limited in their ability to support semantic or percentile-based queries. Examples include identifying which users created the most small files last week or calculating the 99th percentile of directory sizes per group. Executing such queries across a large collection of databases remains inefficient with current monitoring systems. \autoref{tab:admin-queries} lists representative queries that motivate these needs, spanning individual file discovery, anomaly detection, and aggregate usage analysis.

\begin{table*}[ht]
\centering
\caption{Representative queries for file system metadata management.
}
\label{tab:admin-queries}
\begin{tabular}{@{}p{1cm}p{7.2cm}p{6.8cm}p{1.8cm}@{}}
\toprule
\textbf{Role} & \textbf{Example Query} & \textbf{Query Expression} & \textbf{Granularity} \\ \hline
General & Where is a specific file or directory? & \texttt{name LIKE "*pattern*"} & Individual \\ \hline
General & Which files or directories have world-writable permissions? & \texttt{mode = 777} & Individual \\ \hline
General & Which files have not been accessed in over 12 months? & \texttt{atime \textless{} now() - 1y} & Individual \\ \hline
General & Which large files have low access frequency? & \texttt{size \textgreater{} 100GB AND atime \textless{} now() - 6m} & Individual \\ \hline
General & Which files are replicated across directories? & \texttt{GROUP BY checksum HAVING count \textgreater{} 1} & Individual \\ \hline
General & Which directories contain more than 100,000 files? & \texttt{file\_count \textgreater{} 100000} & Aggregate \\ \hline
General & What is the total storage consumption per project? & \texttt{SUM(size) GROUP BY project} & Aggregate \\ \hline
General & Which projects are approaching their quota limits? & \texttt{usage / quota \textgreater{} 0.9} & Aggregate \\ \hline

Admin & Which files are owned by deleted users? & \texttt{uid NOT IN active\_users} & Individual \\ \hline
Admin & Which files exceed their retention period? & \texttt{mtime \textless{} retention\_date} & Individual \\ \hline
Admin & Which users have the most small files? & \texttt{COUNT(file\_size < 1MB) DESC} & Aggregate \\ \hline
Admin & What is the per-user file count and storage usage? & \texttt{SUM(size), COUNT(*) GROUP BY uid} & Aggregate \\ \hline
Admin & What is the 99th percentile directory size per project? & \texttt{PERCENTILE(size, 0.99) GROUP BY project} & Aggregate \\ 

\bottomrule
\end{tabular}
\end{table*}

To address these emerging needs, we present Icicle, a scalable and extensible framework for continuous file system metadata indexing and monitoring across HPC file systems.
Icicle aims to maintain an external, queryable, and unified view of disparate file systems to support semantically rich queries. Icicle operates in two complementary modes: \emph{snapshot mode}, which ingests periodic metadata exports or GUFI-style database dumps; and \emph{update mode}, which continuously consumes and reduces real-time metadata event streams such as Lustre changelogs and GPFS \texttt{mmwatch}~\cite{ibm_mmwatch} events.
By integrating both modes using a modern stream-processing architecture built on Apache Kafka~\cite{kreps2011kafka} and Apache Flink~\cite{carbone2015apache}, Icicle provides scalable, fault-tolerant, and idempotent processing of metadata events. Its open architecture can support any POSIX-compliant environment that exposes metadata event streams or snapshots, and it can directly feed downstream policy engines, dashboards, and workflow systems for real-time automation.

This paper makes three contributions:

\begin{enumerate}
    \item \textbf{The design of a unified and scalable metadata indexing framework.} We introduce Icicle, a scalable metadata indexing and monitoring framework that unifies heterogeneous sources of file system state, including snapshots, GUFI-style databases, and real-time changelogs from Lustre and GPFS. Icicle is open source and available at \url{https://github.com/globus-labs/icicle/}.

    \item \textbf{High-throughput and idempotent stream processing.} Icicle comprises three core components: (i) a high-throughput \emph{ingestion layer} that collects file system metadata from existing tools and services; (ii) a distributed \emph{aggregation and processing layer} built on Kafka and Flink that performs stream-based event reduction and state computation, to transform events into search-engine ingestion requests; and (iii) a \emph{web-based interface} that enforces data visibility for administrators and users to perform queries and custom analytics.

    \item \textbf{Evaluation of scalability and throughput.} We conduct a comprehensive evaluation of Icicle’s snapshot pipeline and event monitor on Lustre and GPFS file systems. In snapshot mode, we show that Icicle scales with both dataset size and available CPU resources, enabling efficient processing of multi-petabyte metadata exports. In update mode, Icicle’s Lustre monitor achieves 4--100$\times$ higher throughput than the state-of-the-art FSMonitor~\cite{paul_fsmonitor_2019}, and both the Lustre and GPFS monitors scale linearly with increasing changelog volumes.

\end{enumerate}

Icicle’s open and extensible design integrates seamlessly with existing HPC tools, establishing a foundation for federated, continuously updated, and queryable views of the file systems. Ultimately, Icicle aims to make data more discoverable, auditable, and manageable regardless of physical location, and advances the next generation of cyberinfrastructure.

\section{Requirements}
\label{sec:requirements}

\autoref{tab:admin-queries} presents representative queries performed by users and administrators across storage environments. These queries span multiple dimensions of file system management, from finding files of interest and identifying anomalies or security concerns to supporting data lifecycle decisions and resource allocation monitoring. Such queries define the \textit{functional} requirements for Icicle's user interfaces. From these, we derive the following \textit{technical} requirements for a monitoring system.

Based on our analysis of operational needs and the limitations of existing tools, we identify the following key requirements for a scalable metadata monitoring framework:

\begin{enumerate}
\item \textbf{Snapshot ingestion.} The system must efficiently ingest metadata exports, including GUFI-style databases and file system snapshots, to provide a consistent baseline view of file system state. Snapshot ingestion enables bulk updates and supports environments where real-time monitoring is unavailable or impractical.

\item \textbf{Real-time monitoring.} The system must continuously process metadata event streams from modern HPC file systems, such as Lustre changelogs and GPFS \texttt{mmwatch} events. 
Real-time processing enables second-level freshness for time-sensitive operations such as anomaly detection and quota enforcement. 

\item \textbf{Configurability.} The system must support flexible trade-offs between consistency, latency, and resource consumption. Anomaly detection may require near-instantaneous updates, while usage trend analysis may tolerate daily or weekly snapshots. The system should accommodate this spectrum of requirements without requiring separate infrastructure for each use case.

\item \textbf{Heterogeneous source support.} The system must support diverse storage backends without requiring significant customization. 
Storing metadata using a common and open index abstraction enables unified querying regardless of the underlying data source. 

\item \textbf{Multi-granular semantic query support.} The system must enable analysis at multiple granularities, ranging from individual files (e.g., locating specific files) to aggregated views reflecting users, groups, or projects. Beyond simple file counting and hierarchical aggregation, the system must support rich semantic queries, including percentile computations, temporal filtering, and cross-attribute correlations.
\end{enumerate}

\section{Icicle}
\label{sec:design}
\begin{figure}[t]
  \centering
  \includegraphics[width=\columnwidth, trim=5.7cm 2.5cm 5.8cm 2cm, clip, page=1]{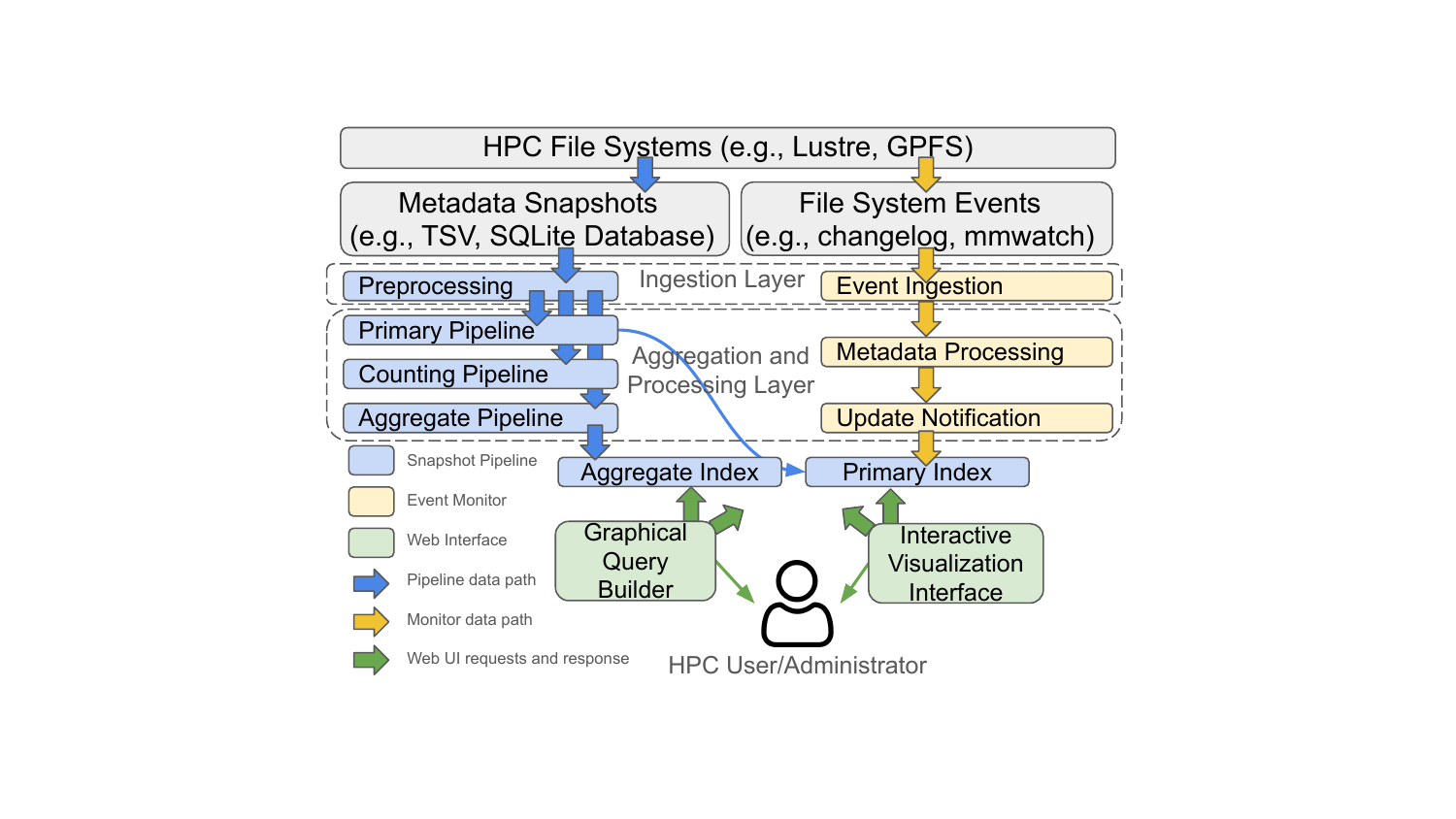}
  \caption{Icicle architecture overview. Icicle comprises a snapshot pipeline, an event monitor, and a web interface. The snapshot pipeline ingests metadata snapshots to populate a per-object primary index and an aggregate index with per-user, group, and directory summaries. The event monitor statefully processes file system events to maintain the primary index. The web interface provides unified access to both indexes.
  }
  \label{fig:architecture}
\end{figure}

Icicle is a scalable file system metadata indexing and monitoring framework. 
We first describe Icicle’s unified metadata model abstraction. We then present its ingestion and distributed aggregation layers, which maintain indexes through both snapshot-based bulk ingestion and event-based real-time synchronization. Finally, we present its web interface, which exposes access to metadata indexes through a unified frontend. \autoref{fig:architecture} shows an overview of the system architecture.

\subsection{Metadata Indexes}\label{sec:indexes}

Icicle organizes file system metadata into two complementary indexes that together provide a comprehensive, queryable view of storage state. 
This dual-index architecture reflects a fundamental trade-off in metadata management: fine-grained, per-object records enable precise queries but incur high storage and query costs, while pre-computed summaries sacrifice granularity for dramatically improved query performance. 
Maintaining both indexes allows Icicle to efficiently support the full spectrum of administrative queries outlined in \autoref{tab:admin-queries}.

The \textbf{primary index} serves as the authoritative record of individual file system objects. Each entry corresponds to a single file or link and captures the core POSIX metadata attributes (see \autoref{tab:primary-index-schema}). 
These records support file-level queries that require object-specific filtering, such as identifying all files owned by a particular user within a directory, locating world-writable files for security auditing, or identifying large files that have not been accessed within a specified period.

The \textbf{aggregate index} provides pre-computed summary statistics for users, groups, and directories. 
Each aggregate record captures statistical summary information for all objects within its scope (see \autoref{tab:aggregate-index-schema}).  
This design supports analytical queries, such as identifying the top storage consumers by project, computing per-user file count distributions, or determining the 99th percentile directory size, without requiring expensive scans over millions or billions of primary records. 

Icicle's metadata model targets POSIX-compliant file systems. Extending support to non-POSIX environments, such as cloud object stores with ACL or IAM-based access control, would require schema changes (e.g., adding IAM principals, bucket identifiers, and storage-class fields) and adapters for cloud-native event streams such as S3 Event Notifications, Google Cloud Pub/Sub, or Azure Event Grid. Icicle's modular architecture accommodates such extensions: new metadata fields require changes only to the preprocessing script and primary pipeline mapper, and new event sources require an additional ingestion-layer adapter.

\begin{table}[t]
\centering
\caption{Primary index schema for file system objects.}
\label{tab:primary-index-schema}
\begin{tabularx}{\columnwidth}{lX}
\toprule
Field & Description \\
\midrule
\texttt{path} &
Fully resolved absolute path; primary key uniquely identifying each object. \\

\texttt{type} &
Object type: regular file (\texttt{f}) or symbolic link (\texttt{l}). \\

\texttt{mode} &
POSIX permission and mode bits (e.g., \texttt{-rw-r--r--}); enables permission-based filtering. \\

\texttt{uid} &
Numeric user ID of the owner; enables per-user queries. \\

\texttt{gid} &
Numeric group ID; supports group-level access and allocation analysis. \\

\texttt{size} &
Object size in bytes; used for storage consumption analysis. \\

\texttt{atime} &
Last access time; identifies cold data and archive candidates. \\

\texttt{ctime} &
Last metadata change time; supports audit and compliance queries. \\

\texttt{mtime} &
Last content modification time; enables age-based lifecycle analysis. \\

\texttt{fileset} &
Containing fileset (GPFS only); supports fileset-scoped queries. \\

\bottomrule
\end{tabularx}
\end{table}
\begin{table}[t]
\centering
\caption{Aggregate index schema for precomputed summary statistics.}
\label{tab:aggregate-index-schema}
\begin{tabularx}{\columnwidth}{lX}
\toprule
Field & Description \\
\midrule
\texttt{principal} &
Aggregation key identifying the scope of the summary: numeric user ID, numeric group ID, or directory path. \\

\texttt{file\_count} &
Total number of files within the aggregation scope; enables rapid object count queries without scanning the primary index. \\

\texttt{size\_\{*\}} &
Distributional statistics over file sizes; supports capacity planning, anomaly detection, and usage reporting. \\

\texttt{atime\_\{*\}} &
Distributional statistics over access times, enabling identification of active versus dormant data. \\

\texttt{ctime\_\{*\}} &
Distributional statistics over metadata change times, supporting audit and compliance queries. \\

\texttt{mtime\_\{*\}} &
Distributional statistics over modification times, enabling age-based lifecycle analysis. \\
\bottomrule
\end{tabularx}

\vspace{0.5ex}
\footnotesize{
\textbf{Note:} \texttt{\{*\}} denotes the set
\{{min}, {p25}, {p50(median)}, {p75}, {max}, {mean}\}; \\
for \texttt{size}, the set additionally includes {total}.
}
\end{table}

\subsection{Data Ingestion, Processing, and Aggregation}

To produce and maintain these indexes, we employ two processing approaches. The \textbf{snapshot pipeline} runs periodically to convert static file system metadata snapshots (e.g., obtained with GUFI indexer or IBM Storage Scale's \texttt{mmapplypolicy}) into the two indexes. The \textbf{event monitor} %operates following the snapshot pipeline to 
subscribes to file system event mechanisms provided by the underlying storage infrastructure to receive continuous updates about file creations, modifications, and deletions. These events are statefully processed into record ingestion and deletion requests that are applied to the primary index.

Icicle’s architecture is designed for extensibility and scale. Both the snapshot pipeline and the event monitor can be applied to \textit{any} storage system that supports snapshots or exposes an event stream. 
To support large deployments, the Icicle monitor scales horizontally by aligning monitor instances with the file system’s metadata partitioning. It deploys one monitor per Lustre MDT or GPFS \texttt{mmwatch} topic so that processing parallelism matches the underlying storage architecture.

\subsubsection{Snapshot pipeline}
The snapshot pipeline provides a scalable workflow for converting HPC metadata snapshots into searchable index structures. Existing metadata indexing tools such as Brindexer, GUFI, and IBM’s \texttt{mmapplypolicy} export file system metadata as per-directory SQLite databases or plain-text listings.
This pipeline %bridges this gap by producing 
produces structured primary and aggregate records tailored for ingestion into search indexes such as Globus Search~\cite{ananthakrishnan2018globus}. 
The snapshot pipeline also embeds version identifiers to support idempotent periodic pipeline execution (e.g., daily or weekly) and to automatically invalidate prior records, allowing users to detect object creation, updates, and deletion over time.

\subsubsection{Event monitor}
The real-time event monitor %supports both Lustre and IBM Storage Scale, 
is designed to maintain index freshness between snapshots by processing file system events as they are emitted. Unlike snapshots, events describe file and directory operations, not state, and thus require stateful reconstruction to derive the final metadata changes. For example, file and directory create, update, and delete events may be canceled or coalesced, and directory rename events propagate recursively to all descendants. Icicle employs a stateful, rule-based reduction model that maintains in-memory directory hierarchy state and applies predefined coalescing and cancellation rules to reduce the event stream to a minimal, semantically equivalent set of metadata updates.

\subsection{Icicle Web Interface}

\begin{figure*}[t]
  \centering
  \setlength{\fboxsep}{0pt} 
  \setlength{\fboxrule}{0pt} 

  \begin{subfigure}[b]{0.338\textwidth} 
    \centering
    \fbox{\includegraphics[
      trim=0cm 0cm 1.8cm 0cm,
      clip,                 
      width=\linewidth, 
      keepaspectratio
    ]{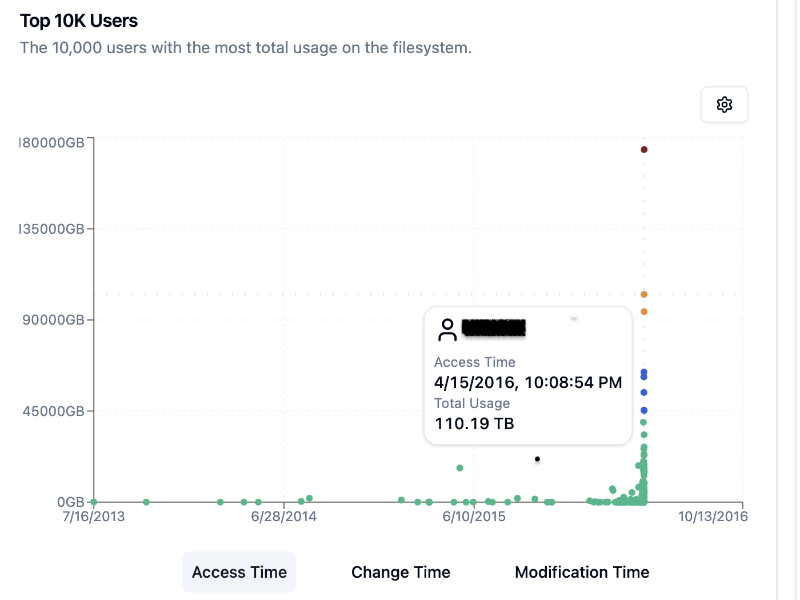}}
    \caption{Top 10K users by storage, rendered from the aggregate index (\autoref{tab:aggregate-index-schema}).}
    \label{fig:top-users}
  \end{subfigure}
  \hfill
  \begin{subfigure}[b]{0.36\textwidth} 
    \centering
    \fbox{\includegraphics[
      trim=0cm 0cm 0cm 0cm,
      clip,
      width=\linewidth, 
      keepaspectratio
    ]{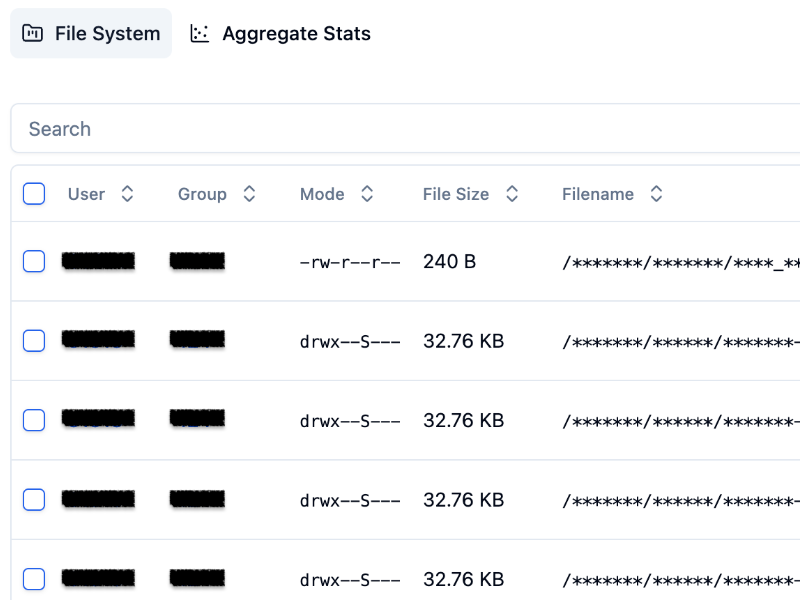}}
    \caption{Graphical query builder issuing queries against the primary index (\autoref{tab:primary-index-schema}).}
    \label{fig:query}
  \end{subfigure}
  \hfill
  \begin{subfigure}[b]{0.285\textwidth}
    \centering
    \fbox{\includegraphics[
      width=\linewidth, 
      keepaspectratio
    ]{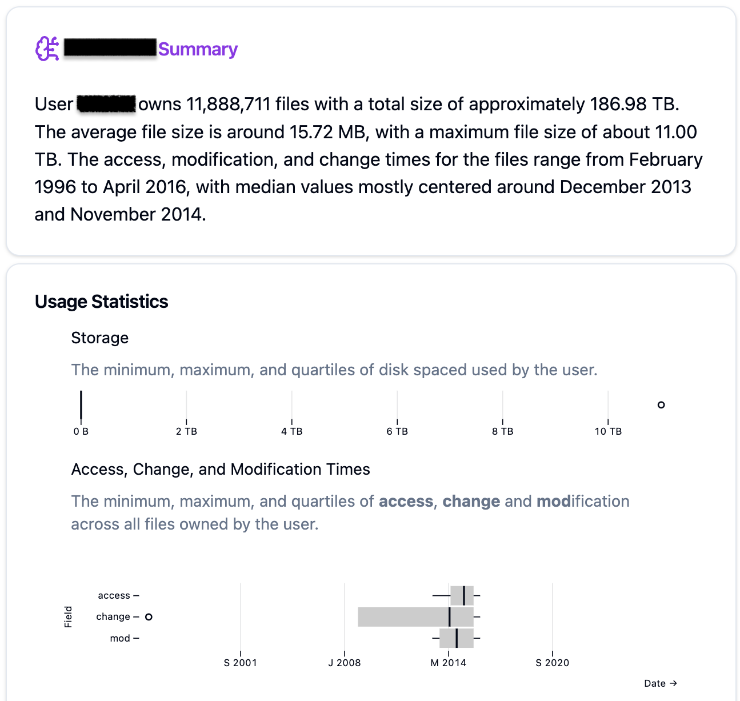}}
    \caption{User summary populated from aggregate index.}
    \label{fig:ai-summary}
  \end{subfigure}

  \caption{Icicle web interface overview.}
  \label{fig:icicle-overview}
\end{figure*}

Icicle provides a web-based interface that unifies access to both primary and aggregate indexes, supporting administrative workflows such as ad hoc querying, capacity planning, and data lifecycle management, as well as user workflows including permission checks and large file or directory discovery. The interface includes a graphical query builder and optional raw query mode supporting regex-match filters on any field; interactive visualizations of storage usage by user, group, and directory; and tools for generating file lists and scheduled reports for policy enforcement and remediation. In addition, the interface also generates summaries by populating structured templates with fields from the aggregate index, enabling rapid, high-level insight into file system behavior at scale. For automated or advanced workflows, both Elasticsearch and Globus Search APIs are directly accessible, enabling programmatic querying, filtering, and aggregation over both indexes. Examples of this web interface are shown in \autoref{fig:icicle-overview}.

\section{Implementation}
\label{sec:implementation}
Here we describe how Icicle's snapshot pipeline and event monitor are implemented. 

\subsection{Pipeline Architecture} 

Icicle’s snapshot pipeline processes file system metadata snapshots and delivers structured records to an external metadata index. Currently, the index is implemented using Globus Search~\cite{ananthakrishnan2018globus}, which is a scalable, secure, and managed indexing and query service for HPC-scale metadata; however, the specific metadata sink is replaceable, allowing use of 
alternative backends such as Elasticsearch~\cite{elasticsearch} and OpenSearch~\cite{opensearch}. % to be integrated easily.

At the ingestion layer, raw metadata snapshots (GUFI SQLite databases or GPFS TSV listings) are preprocessed into compact, uniform CSV files stored in Amazon S3.
Then, at the distributed aggregation and processing layer, the snapshot pipeline operates exclusively on a reduced metadata representation, decoupling file system-specific formats from ingest logic and reducing data volume and I/O overhead.

The snapshot pipeline is implemented using Apache Flink, deployed on Amazon Managed Flink. Managed Flink automatically handles operator retries, state persistence, and dynamic load balancing across Kinesis Processing Units (KPUs), each providing 1 vCPU, 4~GB RAM, and 50~GB ephemeral storage, ensuring high sustained throughput. 
The PyFlink code and its dependencies are packaged as a ZIP archive and stored in Amazon S3, from which workers retrieve them at start time. 

The snapshot pipeline comprises three logically distinct workflows, \textit{primary}, \textit{counting}, and \textit{aggregate}, each expressed as a map–reduce computation. We use Amazon Managed Streaming for Apache Kafka (MSK), via Octopus~\cite{pan_2024_octopus} with Globus Auth integration, for ingest result logging in the primary and aggregate pipelines and for collecting the counting pipeline's results.

\subsubsection{From Preprocessed CSVs to Primary Records}

The \textit{primary} pipeline reads preprocessed CSV files from an S3 bucket and converts each row into a JSON %\texttt{GMeta} record, a JSON-serializable structure 
record suitable for Globus Search ingestion. 
Each %\texttt{GMeta} 
record contains three fields: \texttt{subject}: the file or link path from the file system root; \texttt{visible\_to}: a list of users or groups who will be able to view the indexed record; 
and \texttt{content}: an arbitrary list of key-value pairs describing file attributes such as size, owner ID, group ID, permissions, and access, change, and modification timestamps normalized to ISO8601 with timezone offsets. 
Permission and mode bits are provided both in human-readable form (e.g., \texttt{-rw-r--r--}) and as integers (e.g., 100644) in \texttt{content.raw}. 

During the reduce stage, 
records are accumulated into batches of approximately 10~MB, the maximum payload accepted by the Globus Search ingest API. When either the batch reaches this limit or a 5-second timeout elapses, the accumulated records are 
submitted to the Globus Search ingest API. The asynchronous request IDs returned by the ingest API are published to a dedicated MSK topic for audit logging. Managed Flink executes the primary pipeline in \texttt{streaming} mode, allowing map and reduce operators to run concurrently and to be dynamically redistributed across KPUs, maintaining throughput and resilience.

\subsubsection{Preparation for the Aggregate Pipeline}

The \textit{counting} pipeline computes object counts for each user, group, and directory prefix (truncated to a configurable maximum depth, \texttt{directory\_max}) and prepares these counts as auxiliary input for the aggregate pipeline. Directory counts at this stage are non-recursive; recursive totals are computed in a subsequent post-processing step. For every input CSV row, the map worker assigns an integer shard ID in the range $[0, 63]$ by applying \texttt{zlib.crc32} to the row’s UTF-8 encoding. This hash-based sharding spreads work across workers to scale out processing and reduce skew. It then emits three intermediate tuples of the form (\texttt{principal\_id}, \texttt{shard\_id}, 1) to the reduce workers, where the principal IDs correspond to the owner ID (prefixed with "u"), the group ID (prefixed with "g"), and the directory prefix.

During the reduce stage, all records sharing the same principal ID and shard ID are aggregated into a single (\texttt{principal\_id}, \texttt{shard\_id}, count) record. These outputs are written to a dedicated MSK topic. Once the counting pipeline completes, a script consumes the topic, reconstructs the full directory hierarchy to compute recursive directory counts (because the emitted messages contain only non-recursive counts for directory shards), merges the user and group shard counts, and produces a compact CSV file that is staged alongside the code for use by the secondary pipeline. The counting pipeline operates in \texttt{batch} mode %rather than \texttt{streaming} mode 
to emit one message per (\texttt{principal\_id}, \texttt{shard\_id}) shard.

\subsubsection{From Preprocessed CSVs to Aggregate Records}

The aggregate pipeline computes statistical summaries for each user, group, and directory principal. These summaries include quantiles as well as minimum, maximum, and average values for file sizes and timestamps. It reads preprocessed CSVs from S3 and also consumes the counting file produced by the counting pipeline. %, which specifies the expected number of objects per (\texttt{principal\_id}, \texttt{shard\_id}) pair.

In the map stage, each CSV row is expanded into multiple records, one for each principal associated with that row's path: user, group, and all directory prefixes between \texttt{directory\_min} and \texttt{directory\_max}.
Each emitted tuple has the form (\texttt{principal\_id}, \texttt{shard\_id}, size, atime, ctime, mtime). Map workers maintain mergeable quantile sketches for the four numeric attributes (size, atime, ctime, and mtime), along with running minimums, maximums, and totals. These sketches ensure bounded memory usage and enable scalable quantile estimation, whereas exact quantile computation is expensive in terms of time and resource use. % would exceed the memory limits of Managed Flink for large datasets.
Once all shards for a principal
are received, sketches are serialized and forwarded to the reduce workers.

During the reduce stage, workers merge all sketch shards for a given principal and emit a single 
aggregate record. The \texttt{subject} field encodes the principal (e.g., "user:123", "group:456", or "dir:/a/b/c"), and the \texttt{visible\_to} field mirrors that of primary records. The \texttt{content} field contains quantile estimates ($p10, p25, p50, p75, p90, p99$), minimum and maximum values, total file size, and the file count. 
Flink executes the aggregate pipeline in \texttt{streaming} mode, as with the primary pipeline. However, the reduced volume of aggregate records allows them to be submitted to Globus Search immediately upon creation.

\paragraph{Configurability, Scalability, and Extensibility}

Resource allocation is flexible: Amazon Managed Flink allows each pipeline to scale from 1 to 256 KPUs, and the workflow can be further decomposed along user, group, or directory dimensions (e.g., aggregate only on users or a subset of directories) to accommodate even larger workloads.

Pipeline behavior is highly configurable: the operator can specify the pipeline sinks,
define which users and groups are authorized to view indexed records, and tune batch size thresholds and submission timeouts. The pipeline also allows configuration of aggregation behavior, including the maximum number of shards per aggregation principal and the directory depth for aggregate records.

The system is intentionally modular. Adding new metadata fields requires changes only to the preprocessing script and the primary pipeline’s mapper, while introducing additional aggregate statistics (such as 99.9th percentile) requires modifying a single line in the secondary pipeline’s reduce logic. This modularity allows the pipeline to evolve alongside file system features and analysis requirements. % while preserving its overall architecture.

\subsection{Monitor Architecture}

The monitor consists of three layers: event ingestion, metadata processing, and update notification. The ingestion layer unifies %Lustre changelogs and GPFS inotify-style 
events into an internal format. The metadata-processing layer applies stateful reduction rules. % and reconstructs file paths using a lightweight, in-memory model of the directory hierarchy. 
The update notification layer emits path-resolved update and deletion events via MSK or directly to Globus Search. The design supports extensible reduction rules, pluggable storage models, and tunable retention windows for directory hierarchy tracking.

\subsubsection{Event Ingestion Layer}

The ingestion layer abstracts the heterogeneous event formats of Lustre and GPFS into uniformly structured key-value pairs containing the fields required for stateful reconstruction of metadata updates.

For Lustre, the monitor invokes Lustre utilities (\texttt{lfs changelog} and \texttt{lfs changelog\_clear}) via subprocess calls. It parses the raw textual output into structured dictionaries containing event ID, event type, timestamps, and file identifiers (FIDs). Different Lustre event types require different parsing logic; for example, creation and deletion events include a \texttt{parent\_fid}, whereas rename events contain \texttt{source\_fid} and \texttt{source\_parent\_fid}. A key performance optimization is to avoid \texttt{lfs fid2path} at this stage. Since this call incurs $\sim$10 ms of latency per invocation, path resolution is deferred to the metadata-processing layer, where it is invoked only when necessary (i.e., the changelog is not eliminated after applying stateful reduction rules).

For GPFS, a \texttt{mmwatch} watcher can be configured to send fileset changelogs to a Kafka topic. The monitor can spawn one or more \texttt{confluent\_kafka}~\cite{confluent_kafka_python} consumers operating in parallel, each assigned to different topic partitions. Events are serialized with the \texttt{orjson}~\cite{orjson} library and transported to the metadata-processing layer via a high-throughput, multiple-producer-single-consumer in-memory queue.

The monitor supports optional ingestion-layer filtering to discard high-volume, low-information events (e.g., file open events), reducing metadata processing %layer processing
overhead.

\subsubsection{Metadata Processing Layer}

The metadata processing layer applies reduction rules to statefully transform event streams into a minimal, semantically equivalent set of metadata updates. 
We include three types of reduction rules: 
(1) {update coalescing}: multiple %metadata or data change 
events for the same FID are reduced to a single event, as a subsequent \texttt{stat} call can capture the final state of the object; 
(2) {event cancellation}: transient operation sequences, such as a \texttt{CREAT} followed by a \texttt{UNLNK} for the same file, or \texttt{MKDIR} followed by a \texttt{UMDIR} for the same directory, within the same batch are eliminated;
(3) {rename override}: the above reductions are bypassed when a directory rename event occurs, as moving a directory affects both its original and destination parents as well as all descendant objects.

Events that pass these reductions are stored in FID-keyed slots until a batch is ready. Batching is triggered either by reaching a size limit (e.g., 1000 events) or a time threshold (e.g., 5 seconds of inactivity), balancing throughput and update latency. The batch is then forwarded to the state manager (in the metadata processing layer), which maintains an in-memory representation of the file system hierarchy. 
Using parent–child relationships, the state manager resolves operations such as path construction for newly created objects without invoking \texttt{lfs fid2path} on Lustre, and recursively updates descendant paths affected by directory rename operations. 
Processing logic is dispatched to file system-specific handlers for creation, deletion, update, and rename events for both Lustre and GPFS.

The state manager emits two lists: (1) \texttt{to\_update}: with FID, path, and stat for each file and link object, where stat is gathered via \texttt{stat} calls on Lustre but directly carried from GPFS \texttt{mmwatch} changelogs. (2) \texttt{to\_delete}: containing the FID and resolved path of objects whose primary records must be removed or invalidated in the index. 

\subsubsection{Update Notification Layer}

The final stage converts the two lists from the state manager into Globus Search ingestion and deletion requests. Depending on deployment configuration, this layer can issue requests directly to Globus Search or publish them to a dedicated MSK topic for further inspection and integration with downstream consumers, such as user notification services. This design supports both synchronous ingestion for low-latency metadata updates and asynchronous processing for monitoring, auditing, or additional analytics.

\subsubsection{Scalability and Persistence}

To determine which objects are affected by directory rename events, each Lustre monitor maintains an in-memory directory hierarchy for the MDT from which events originate. Similarly, in GPFS, the monitor must consume all events from a given fileset topic to ensure correct path reconstruction. 
The monitor can be dedicated to a single MDT in Lustre or to a single Kafka topic for an individual GPFS fileset, allowing monitor deployment to scale linearly with the number of metadata servers and filesets. When resource sharing is desirable, a single monitor instance may watch multiple MDTs or multiple Kafka topics corresponding to different filesets.
For both Lustre and GPFS, an optional LRU-based eviction policy limits the retention of inactive directory entries, reducing memory footprint while preserving correctness for active paths. Together, these mechanisms enable the monitor to scale with file system size and workload intensity while keeping memory usage bounded.

\section{Evaluation}
\label{sec:evaluation}
\begin{table*}[t]
\centering
\caption{File System Dataset Statistics.}
\label{tab:dataset-stats}

\setlength{\tabcolsep}{6pt}
\renewcommand{\arraystretch}{1.1}

\begin{tabular}{l r l r r r r r r r}
\toprule
\multicolumn{2}{c}{File system} &
\multicolumn{4}{c}{Raw Metadata} &
\multicolumn{4}{c}{Preprocessed Metadata} \\
\cmidrule(lr){1-2}
\cmidrule(lr){3-6}
\cmidrule(lr){7-10}

Name & Size &
Source & Size & \# Rows & \# Cols &
\# CSVs & Size & \# Rows & \# Cols \\
\midrule

FS-small &
\num{67.63}~TB &
GUFI \texttt{entries} table &
\num{90.75}~GB & \num{8.46}\text{M} & \num{22} &
\num{9} & \num{1.34}~GB & \num{8.46}\text{M} & \num{9} \\

FS-medium &
\num{1.55}~PB &
\texttt{mmapplypolicy LIST} &
\num{38.83}~GB & \num{145.59}\text{M} & \num{18} &
\num{129} & \num{21.68}~GB & \num{128.50}\text{M} & \num{10} \\

FS-large &
\num{53.59}~PB &
GUFI \texttt{entries} table &
\num{2.15}~TB & \num{1.04}\text{B} & \num{22} &
\num{994} & \num{194.63}~GB & \num{1.04}\text{B} & \num{9} \\

\bottomrule
\end{tabular}
\end{table*}

We evaluate the performance of the pipeline and the monitor to assess:
(1) scalability of the pipeline;
(2) effectiveness of the approximation algorithm;
(3) scalability of the monitor on Lustre as the number of MDTs increases; and
(4) scalability of the monitor on GPFS as the number of filesets increases.

\subsection{Pipeline Evaluation}

\subsubsection{Datasets} 
We evaluate the snapshot pipeline using three real-world HPC file system metadata snapshots: FS-small, FS-medium and FS-large.
FS-small and FS-large are GUFI snapshots from project file systems at a large research computing center. % as collections of per-directory SQLite databases. 
FS-small is a partial index of a GPFS file system, covering the subtree that hosts organizational and staff home directories; it spans 67.63~TB of data and consists of 1.77 million GUFI \textit{entries} tables with 8.46 million rows in total. 
FS-large is from an HPSS tape archive representing 53.59~PB of data across 40.88 million \textit{entries} tables and 1.04 billion rows. Each \textit{entries} row contains 22 metadata fields. %, though some fields may be null. 
FS-medium is from a national supercomputing facility and captures a full production GPFS. It comprises 145.59 million TSV records with 18 metadata fields, describing all files, links, and directories in a 1.55~PB file system. Important characteristics of these datasets are shown in \autoref{tab:dataset-stats}.

\subsubsection{Preprocessing}

The preprocessing stage converts the input SQLite databases and TSV listings into CSV files that contain only the attributes required for the primary and aggregate indexes. 
File paths are escaped to safely handle special characters and ensure robust parsing.
CSV files are generated with a target size of approximately one million rows; however, for FS-small and FS-large, individual files may exceed this threshold since splitting decisions are applied only after fully processing a GUFI database. 
The resulting CSVs are uploaded to a dedicated S3 bucket for each file system and consumed in parallel by the Flink pipeline. Preprocessing produced 9 CSV files (1.34~GB) for FS-small, 129 files (21.68 GB) for FS-medium, and 994 files (194.63~GB) for FS-large.

Metadata preprocessing substantially reduces data volume by over 90\% for FS-small and FS-large, and by over 40\% for FS-medium. 
For GUFI-based snapshots, this reduction is achieved by retaining only ingestion-relevant attributes from the \textit{entries} table and consolidating per-directory SQLite databases into million-row, headerless CSV files. 
For FS-medium, the total row count is reduced from 145.59 M to 128.50 M by filtering out directory entries and retaining only files and links.

\subsubsection{Pipeline Runtime}

\begin{table}[t]
\centering
\caption{Per-stage runtimes (seconds) and normalized total runtime per file system (normalized to 128 KPU = 1).}
\label{tab:pipeline-runtime}
\setlength{\tabcolsep}{5pt}
\begin{tabularx}{\linewidth}{ll *{4}{>{\raggedleft\arraybackslash}X}}
\toprule
\multirow{2}{*}{File system} &
\multirow{2}{*}{KPU} &
\multicolumn{1}{c}{Primary} &
\multicolumn{1}{c}{Counting} &
\multicolumn{1}{c}{Aggregate} &
\multicolumn{1}{c}{Normalized} \\
& &
\multicolumn{1}{c}{Pipeline} &
\multicolumn{1}{c}{Pipeline} &
\multicolumn{1}{c}{Pipeline} &
\multicolumn{1}{c}{Total} \\
\midrule

FS-small        & 128 & \num{79.82}  & \num{166.88} & \num{263.11} & \num{1.00} \\
FS-small        & 256 & \num{74.55}  & \num{107.64} & \num{242.95} & \num{0.83} \\
FS-small$^{*}$  & 128 & \num{62.69}  & \num{116.48} & \num{97.56}  & \num{0.54} \\
\addlinespace

FS-medium       & 128 & \num{132.63} & \num{354.01} & \num{1262.63} & \num{1.00} \\
FS-medium       & 256 & \num{118.87} & \num{337.54} & \num{888.68}  & \num{0.77} \\
\addlinespace

FS-large        & 128 & \num{926.83} & \num{3554.24} & \num{8552.87} & \num{1.00} \\
FS-large        & 256 & \num{483.72} & \num{1832.08} & \num{4820.03} & \num{0.55} \\

\bottomrule
\end{tabularx}

\vspace{0.5ex}
\footnotesize{
$^{*}$Uses a finer-grained CSV partitioning with a target of 100K rows per file.
}
\end{table}

In \autoref{tab:pipeline-runtime}, we report the runtime of each pipeline workflow across three file system datasets and two KPU configurations. Across all file systems, the aggregate pipeline consistently takes longer than the primary pipeline because it requires cross-KPU shuffles to aggregate records by users, groups, and directories, whereas the primary pipeline performs local, in-place aggregation within each KPU, emitting fixed-size (10 MB) bundles without shuffle.

With 128 KPUs, the total execution time of all three pipelines is approximately 8 minutes for FS-small, 29 minutes for FS-medium, and 217 minutes for FS-large. Increasing the number of KPUs yields a limited speedup for FS-small and FS-medium, as their inputs are preprocessed into only 9 and 129 CSV files, respectively. Because input files are assigned at the file granularity, additional KPUs cannot further parallelize ingestion, leaving the S3 read stage the dominant bottleneck. 
In contrast, FS-large contains 994 CSV files, and scaling from 128 to 256 KPUs reduces runtime by approximately 45\%. 
These results show that preprocessing is a key scalability enabler: chunking metadata snapshots into smaller files is essential to fully utilize Flink parallelism and avoid underutilized KPUs. 
To further examine the impact of input granularity, we re-chunk the FS-small dataset into 100K-row CSV files (85 files total). This change improves performance across all pipelines, reducing overall execution time by 46\% and aggregate pipeline runtime by 37\%. % of its baseline.
The aggregate pipeline benefits most from finer partitioning because it runs in streaming mode, where map, shuffle, and reduce execute concurrently; increasing input files from 9 to 85 reduces its runtime from 263\,s to 97\,s. The counting pipeline runs in batch mode with barrier synchronization between stages, limiting its speedup. This difference explains why the counting pipeline runtime exceeds the aggregate pipeline runtime for FS-small*.

The resulting index statistics are reported in \autoref{tab:index-stats}. The primary index size scales with the number of records. FS-medium contains many more directories because we index two levels into user home directories (vs. one level for the others). In all cases, the aggregate index is under 1~GB, making it easy for the web interface to search and query.

\begin{table*}[t]
\centering
\caption{Primary and aggregate index statistics.}
\label{tab:index-stats}

\setlength{\tabcolsep}{6pt}

\begin{tabular}{l r r r r l r r}
\toprule
\multirow{2}{*}{File system} &
\multicolumn{2}{c}{Primary Index} &
\multicolumn{5}{c}{Aggregate Index} \\
\cmidrule(lr){2-3} \cmidrule(lr){4-8}
& \# Records & Size & \# Users & \# Groups & Directory Depth & \# Directories & Size \\
\midrule

FS-small  &
\num{8.46}M &
\num{5.78}~GB &
\num{37} &
\num{12} &
One level into user home directories &
\num{1133} &
\num{3.48}~MB \\

FS-medium &
\num{128.50}M &
\num{90.85}~GB &
\num{240} &
\num{178} &
Two levels into user home directories &
\num{65190} &
\num{193.17}~MB \\

FS-large  &
\num{1.04}B &
\num{774.83}~GB &
\num{2091} &
\num{325} &
One level into user home directories &
\num{16724} &
\num{56.95}~MB \\

\bottomrule
\end{tabular}
\end{table*}

\subsubsection{Approximation Algorithms}

We evaluate the aggregate pipeline's quantile approximation accuracy by comparing the sketch estimates ($\hat{q}$) against exact ground-truth quantiles ($q_{\text{exact}}$) across six target percentiles ($p10$--$p99$). The four sketching algorithms we chose are DDSketch~\cite{masson2019ddsketch}, KLLSketch~\cite{karnin2016optimal}, ReqSketch~\cite{cormode2023relative}, and t-Digest~\cite{dunning2021tdigest}, drawn from the Datadog and Apache DataSketches libraries~\cite{datadog_sketches_py,datasketches_python}. We use their default relative-error guarantees.

For each user and group, and for each distributional attribute of the aggregate index ({size}, {atime}, {ctime}, and {mtime}), we evaluate sketch accuracy using two error metrics over an aggregation of size $N$ (the number of files): \emph{relative value error} $|\hat{q}-q_{\text{exact}}| / |q_{\text{exact}}|$ and \emph{normalized rank error} $|\hat{r}-r^\ast| / N$. Here, $\hat{r}$ is the position of the estimated quantile $\hat{q}$ in the exact sorted list, and $r^\ast$ is the expected rank of the target quantile (e.g., $0.5N$ for the median). The aggregate pipeline using these sketches exhibits comparable runtime across configurations, with DDSketch completing slightly faster (see the aggregate pipeline runtime column of \autoref{tab:sketch-error-summary}).

We observe a fundamental trade-off between rank and value accuracy across sketches. DDSketch provides the most stable value estimates, maintaining a mean relative error below $0.01$ across all file systems, but at the cost of higher normalized rank error (worst-case mean $=0.31$). In contrast, KLLSketch, ReqSketch, and t-Digest achieve superior rank accuracy (worst-case mean $<0.11$) but exhibit large value errors, particularly around the median. This behavior stems from their internal bias toward accurately representing distribution tails, which leaves the central quantiles under-resolved and leads to large relative deviations in high-cardinality datasets such as FS-large. Given that our aggregate pipeline prioritizes value accuracy, we adopt DDSketch as the default sketch algorithm. 

Although exact aggregation is faster on FS-small (71\,s vs.\ 97–98\,s; \autoref{tab:sketch-error-summary}), it requires reduce workers to hold complete value distributions in memory. For FS-medium and FS-large, this exceeds per-KPU memory capacity, as principals such as the root directory must aggregate across all files. DDSketch avoids this limitation with fixed-size summaries and a mean relative value error below 0.01, making sketch-based aggregation preferable for all production deployments.

\begin{table*}[t]
\centering
\caption{
Summary of sketch error across file systems for users and groups with at least 100 files. Min$_q$ and Max$_q$ report errors over the six quantiles $p10$–$p99$, averaged across runs and aggregation keys. The aggregate pipeline runtime column reports the time to compute user and group aggregation (excluding directories), not live query retrieval; once built, the aggregate index supports sub-two-second query response times. Lower is better (↓). 
}
\label{tab:sketch-error-summary}
\setlength{\tabcolsep}{6pt}
\begin{tabular}{llrcccc}
\toprule
\multirow{2}{*}{File system} &
\multirow{2}{*}{\begin{tabular}{c}
Algorithm
\end{tabular}} &
\multirow{2}{*}{\begin{tabular}{c}
Aggregate Pipeline \\ Runtime (s)
\end{tabular}} &
\multicolumn{2}{c}{Mean Normalized Rank Error (↓)} &
\multicolumn{2}{c}{Mean Relative Value Error (↓)} \\
\cmidrule(lr){4-5} \cmidrule(lr){6-7}
& & & Min$_q$ & Max$_q$ & Min$_q$ & Max$_q$ \\
\midrule
\multirow{5}{*}{FS-small}
& Exact     & 71  & --     & --     & --     & --     \\
& DDSketch & 97  & \textbf{0.1023} & \textbf{0.2620} & 0.0048 & 0.0057 \\
& KLLSketch      & 98  & 0.0686 & 0.1097 & 0.0027 & \textbf{0.0666} \\
& ReqSketch& 98  & 0.0705 & 0.1095 & 0.0017 & 0.0317 \\
& t-Digest & 98  & 0.0708 & 0.1124 & \textbf{0.0096} & 0.0216 \\
\midrule
\multirow{5}{*}{FS-medium}
& Exact     & 415 & --     & --     & --     & --     \\
& DDSketch & 347 & \textbf{0.1493} & \textbf{0.3110} & 0.0066 & 0.0080 \\
& KLLSketch      & 350 & 0.0281 & 0.0683 & 0.0027 & 0.1172 \\
& ReqSketch& 356 & 0.0276 & 0.0693 & 0.0040 & 0.0372 \\
& t-Digest & 349 & 0.0277 & 0.0733 & \textbf{0.0116} & \textbf{0.3747} \\
\midrule
\multirow{5}{*}{FS-large}
& Exact     & 4,993 & --     & --     & --     & --     \\
& DDSketch & 2,478 & \textbf{0.1837} & \textbf{0.2628} & 0.0051 & 0.0058 \\
& KLLSketch      & 2,526 & 0.0189 & 0.0406 & 0.0056 & 0.1629 \\
& ReqSketch& 2,607 & 0.0182 & 0.0414 & 0.0058 & \textbf{2.2493} \\
& t-Digest & 2,554 & 0.0182 & 0.0456 & \textbf{0.0135} & 1.9477 \\
\bottomrule
\end{tabular}
\end{table*}

\subsection{Monitor Evaluation}

We deploy Lustre and GPFS file systems on AWS to evaluate the Icicle monitor under backlogged metadata workloads.

\subsubsection{Lustre setup}

All Lustre experiments were conducted on an EC2–based Lustre cluster. The cluster comprised one Management Server (MGS) and two Object Storage Servers (OSSs) hosting four Object Storage Targets (OSTs), providing 1 TB of data storage. The number of OSSs and the available data capacity are not performance bottlenecks in our evaluation, since Icicle interacts exclusively with MDTs and does not issue data-path I/O to OSTs. Across different experiments, we configured 1, 2, or 4 Metadata Targets (MDTs) on 1, 2, or 4 Metadata Servers (MDSs). Each MDS was backed by a 32 GB \texttt{gp3} EBS volume using the default baseline performance of 3000 IOPS and 125 MiB/s throughput. All servers were deployed on \texttt{c5a.large} instances in \texttt{us-east-1a}, while client nodes ran in \texttt{us-east-1d}, with an average server–client RTT of 0.757\,ms. All servers run RHEL 8, while clients run Ubuntu 24.04. Throughout this section, ``FSMonitor" denotes an Icicle baseline that uses
FSMonitor Algorithm~1\cite{paul_fsmonitor_2019}  for FID resolution, while leaving other components of the monitor unchanged; both baselines emit metadata update/delete requests to MSK (unlike the original FSMonitor, which outputs path-resolved changelogs rather than metadata changes).

\subsubsection{Lustre baseline}

We first evaluate Icicle monitor on a single-MDT cluster using the two workloads used to evaluate FSMonitor~\cite{paul_fsmonitor_2019}: evaluate output script (\texttt{eval\_out}) and evaluate performance script (\texttt{eval\_perf}). \texttt{eval\_out} repeatedly exercises a sequence of metadata operations within a directory: each iteration creates a uniquely named file, appends to it, renames it, creates a directory, moves the renamed file into the directory, and then recursively deletes the directory. In contrast, \texttt{eval\_perf} stresses metadata throughput further by repeatedly performing a create–modify–delete cycle on uniquely named files, producing changelog events dominated by file create, opens, closes, and unlinks.

\begin{table}[t]
\centering
\setlength{\tabcolsep}{3pt}
\caption{Average throughput (changelogs/s) for one client and one MDT with workloads \texttt{eval\_out} (evaluate output) and \texttt{eval\_perf} (evaluate performance). \textbf{Chg}: Icicle receives and emits changelogs without stateful processing; \textbf{FSMonitor}: Icicle with FSMonitor-style FID resolution; \textbf{Icicle}: Icicle baseline; \textbf{Icicle+Red.}: Icicle with changelog reduction. \texttt{c5a.large} (2\,vCPUs, 4\,GiB RAM) and \texttt{c5a.xlarge} (4\,vCPUs, 8\,GiB RAM), each with AMD EPYC processor clocked up to 3.3 GHz. Each successive size (xlarge, 2xlarge, 4xlarge, \ldots) doubles the number of vCPUs and memory.}
\label{tab:lustre-base}
\begin{tabular}{ll
                S[table-format=5.0]
                S[table-format=3.0]
                S[table-format=5.0]
                S[table-format=5.0]}
\toprule
\textbf{Client} & \textbf{Workload} &
\multicolumn{1}{r}{\textbf{Chg}} &
\multicolumn{1}{r}{\textbf{FSMonitor}} &
\multicolumn{1}{r}{\textbf{Icicle}} &
\multicolumn{1}{r}{\textbf{Icicle+Red.}} \\
\midrule

\texttt{c5a.large}  & \texttt{eval\_out}  & 34786 & 554 & 32162 & 35680 \\
\texttt{c5a.large}  & \texttt{eval\_perf} & 35089 & 391 & 32553 & 40471 \\
\texttt{c5a.xlarge} & \texttt{eval\_out}  & 35434 & 565 & 32678 & 36194 \\
\texttt{c5a.xlarge} & \texttt{eval\_perf} & 35843 & 406 & 33179 & 41120 \\
\bottomrule
\end{tabular}
\end{table}

\autoref{tab:lustre-base} compares Icicle monitor and FSMonitor changelog processing throughput on a single-MDT Lustre cluster under the two workloads. Across all configurations, FSMonitor achieves substantially lower throughput (391–565 changelogs/s), reflecting the high cost of synchronous \texttt{fid2path} resolution ($\sim$10\,ms per call) performed for every changelog. In contrast, Icicle throughput is comparable to raw changelog ingestion, processing 32–33K changelogs/s, a 57–83$\times$ improvement over FSMonitor. This improvement stems from avoiding per-event \texttt{fid2path}: Icicle resolves an experiment's directory FID once and then constructs descendant paths using parent-child directory state.

Enabling changelog reduction further improves performance, particularly for the \texttt{eval\_perf} workload, where Icicle with reduction reaches up to 41K changelogs/s, exceeding both Icicle baseline and Icicle receiving and emitting changelogs without stateful processing. These gains demonstrate that eliminating highly frequent but low-value \texttt{10OPEN}s (we keep \texttt{11CLOSE}s) and \texttt{01CREAT}/\texttt{06UNLNK} event pairs in the batch processor before they reach the state manager effectively reduces downstream processing overhead, yielding a consistent 1.1–1.2$\times$ throughput improvement over the base monitor configuration.

\autoref{tab:lustre-base} shows that increasing the client instance size from \texttt{c5a.large} to \texttt{c5a.xlarge} results in only modest throughput gains ($\sim$2\%), confirming that the single-MDT configuration, rather than client-side CPU or memory resources, is the dominant bottleneck in this setting.

\subsubsection{Lustre scaling}

\begin{figure}[t]
  \centering
  \includegraphics[width=\columnwidth]{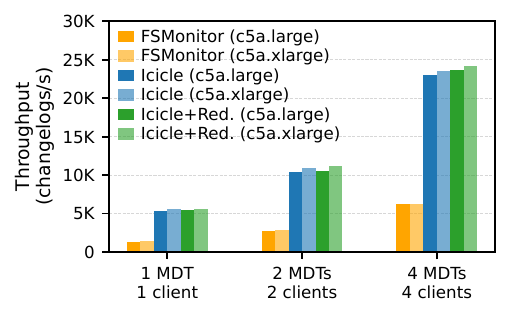}
  \caption{Filebench throughput scaling on Lustre as the MDT count increases with \texttt{c5a.large} or \texttt{c5a.xlarge} clients, with one client per MDT.}
  \label{fig:lustre-scaling}
\end{figure}

We evaluate the monitor’s scaling performance using a Filebench~\cite{tarasov_filebench_2016} workload that generates realistic, metadata-intensive access patterns. % and was also used to evaluate FSMonitor. 
The workload pre-populates a directory tree with 50K files whose sizes follow a Gamma distribution (mean size $\sim$16~KB, $\gamma=1.5$), organized with an average directory width of 20 and a mean directory depth of 3.6. During execution, 32 concurrent threads repeatedly perform open–read–close operations on randomly selected files for 180\,s, producing a sustained stream of fine-grained metadata events.

\autoref{fig:lustre-scaling} shows throughput scaling using \texttt{c5a.large} and \texttt{c5a.xlarge} clients as the number of MDTs increases. Icicle scales nearly linearly, from $\sim$5.4K changelogs/s on one MDT to over 23K changelogs/s on four MDTs. Changelog reduction yields only modest gains (0–2\%), as the Filebench workload lacks create–delete patterns, limiting reduction to filtering \texttt{10OPEN} events before they reach the state manager. Placing monitors on \texttt{c5a.xlarge} yields only modest gains (2–7\%), suggesting throughput is limited mainly by MDT parallelism and monitor event processing, not client resources.

On the Filebench workload, FSMonitor achieves higher throughput in the single-MDT configuration ($\sim$1.36K changelogs/s) than on the \texttt{eval\_out} and \texttt{eval\_perf} workloads (0.3--0.5K changelogs/s). This is because Filebench does not delete files after the initialization phase, allowing FSMonitor to reuse cached \texttt{fid2path} resolutions for repeated open, read, and close operations, thereby reducing lookup overhead. In contrast, although Icicle performs only a single \texttt{fid2path} resolution at the experiment directory root and derives all subsequent paths recursively from parent directory state, it exhibits lower throughput in this setting than in the previous two workloads in the baseline evaluation, because it performs additional per-event bookkeeping to maintain directory states, which may invoke per-file \texttt{stat}. Nevertheless, Icicle still achieves 3.68–4.05$\times$ higher throughput than FSMonitor.

\subsubsection{GPFS Scaling}

\begin{figure}[t]
  \centering
  \includegraphics[width=\linewidth]{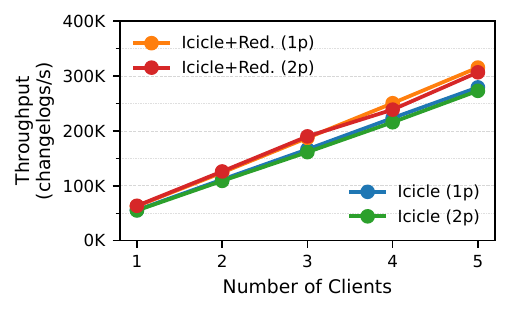}
  \caption{
    Filebench throughput scaling on GPFS as a function of the number of clients, with one \texttt{c5a.large} client per fileset. \texttt{1p} denotes a monitor consuming from one Kafka partition, while \texttt{2p} uses two partitions merged into a single state manager. The limited gain from \texttt{2p} indicates that throughput is constrained by fileset parallelism and event processing.
  }
  \label{fig:gpfs-scaling-clients}
\end{figure}

\begin{figure}[t]
  \centering
  \includegraphics[width=\linewidth]{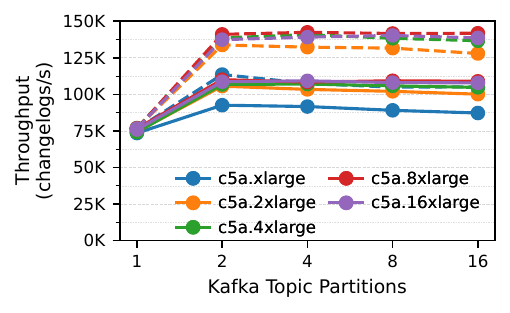}
  \caption{
    Throughput versus Kafka topic partitions for a single GPFS client (one fileset). Solid lines denote Icicle, and dashed lines denote Icicle with event reduction. Throughput saturates beyond two partitions due to aggregation at the state manager.
  }
  \label{fig:gpfs-scaling-partition-icicle}
\end{figure}

We evaluate Icicle’s scalability on GPFS by configuring a single-node cluster running on \texttt{c5a.xlarge} to stream changelog events directly to a local Kafka broker. We apply the same Filebench workload used in the Lustre experiments to each fileset. 
We vary the number of (1) watched filesets (1--5); (2) client nodes (one \texttt{c5a.large} client per fileset); and (3) Kafka topic partitions (1 or 2). To explore upper bounds, we additionally scale client instances up to \texttt{c5a.16xlarge}, increase Kafka topic partitions up to 16, and scale the GPFS+Kafka server to \texttt{c5a.16xlarge}. Servers and clients are set up in \texttt{us-east-1d}, and the round-trip latency is $\sim$0.487\,ms. % \ryan{XX}

\autoref{fig:gpfs-scaling-clients} shows that Icicle scales linearly with the number of clients. This setup mirrors the Lustre experiments, where we add one client per MDT; here, we add one client per fileset. Without event reduction, throughput reaches 55.7K changelogs/s for a single fileset and increases to 279.5K changelogs/s for five filesets. With event reduction enabled, throughput improves from 62.9K to 315.7K changelogs/s, maintaining linear scaling even when both GPFS and Kafka run on a single \texttt{c5a.xlarge} server.

We also evaluate configurations where each client runs two or more Kafka consumers (i.e., consuming a multi-partition topic), as shown in \autoref{fig:gpfs-scaling-clients} for \texttt{c5a.large} clients and two-partition topics and \autoref{fig:gpfs-scaling-partition-icicle} for \texttt{c5a.xlarge} clients and multi-partition topics. For \texttt{c5a.large} clients, performance is comparable to the single-partition configuration and is occasionally slightly lower, indicating limited benefit from additional partitions at this scale. In contrast, scaling up client resources has a clear impact: a \texttt{c5a.xlarge} client achieves 92.6K changelogs/s with a two-partition topic, compared to 55.0K changelogs/s for a \texttt{c5a.large} client, indicating that the client CPU is a primary constraint.

Further scaling beyond \texttt{c5a.2xlarge} clients or more than four partitions yields diminishing returns, as performance becomes limited by the state manager’s ability to process GPFS changelogs. Notably, even the single-partition GPFS configuration significantly outperforms Lustre. This advantage stems from GPFS inotify-based changelogs, which include file metadata (e.g., \texttt{stat} information) directly in the event stream. As a result, Icicle avoids per-file \texttt{stat} calls in the state manager, substantially reducing processing overhead.

To determine whether these findings hold beyond single-node deployments, we also evaluated two-node and four-node GPFS configurations. Performance remained bounded by Kafka consumption throughput rather than event generation.
Unlike Lustre, where adding MDTs introduces independent metadata partitions that the monitor can consume in parallel, GPFS scaling is constrained by the rate at which the monitor processes events from Kafka topics. Increasing the server instance size similarly provides minimal throughput improvement, confirming that the state manager's processing capacity, not deployment topology or server-side resources, is the primary throughput bottleneck for GPFS.

\section{Related Work}
\label{sec:related}

\textbf{HPC File System Metadata Indexing.}
A large body of work has focused on improving the performance and scalability of metadata indexing and search in HPC file systems. Early efforts primarily employ tree-based structures such as k-d trees %~\cite{friedman1975algorithm}
and R-trees %~\cite{guttman1984r}
to accelerate metadata queries over hierarchical namespaces~\cite{leung_spyglass_2009, hua_smartstore_2009, parker-wood_security_2010, patil_scale_2011}. TableFS~\cite{ren2013tablefs} adopts an LSM-tree-based design to optimize metadata insertion throughput. Subsequent distributed file system metadata planes~\cite{ren2014indexfs, zheng2015deltafs, lv2022infinifs} improve the scalability of metadata operations such as file creation and deletion.

External metadata indexing systems decouple query performance from file system operations while reducing memory pressure~\cite{cipar_lazybase_2012, borgfs_2014}. Brindexer~\cite{paul_efficient_2020} parallelizes namespace scans into SQLite databases, enabling efficient queries via RDBMS-style partitioning and multi-threaded traversal; however, it still relies on static snapshots. % and offers only limited support for processing Lustre changelogs in real time. 
GUFI~\cite{manno_gufi_2022} constructs a hierarchy of per-directory SQLite databases that preserve POSIX permissions and enable interactive metadata queries. 
Despite these improvements, GUFI’s architecture remains scan-based, requiring periodic rebuilds of read-only snapshots to maintain consistency with the underlying file system.

The Robinhood Policy Engine~\cite{leibovici_taking_2015} is another widely deployed external metadata indexing system. It mirrors system metadata in a local database (via scans or changelogs) and enforces policies for purge, hierarchical storage management (HSM), and OST-balancing. While its scan mode supports other POSIX-compliant systems, Robinhood mainly supports Lustre by polling changelogs from Lustre metadata servers and storing metadata (inferred from changelogs and \texttt{stat} calls) in a single SQL database, which limits scalability and flexibility.

Recent work decomposes policy enforcement into event-driven agents. For example, QuickSilver~\cite{brumgard_quicksilver_2022} and PoliMOR~\cite{george_polimor_2023} organize lifecycle management into scan, policy, and action agents that communicate over message queues with minimal shared state. Similarly, commercial engines, such as Cray ClusterStor Data Services~\cite{hpe_Cray_ClusterStor} and GPFS Information Lifecycle Management~\cite{ibm_spectrum_scale_ilm}, provide deep product integration, but they remain vendor-specific~\cite{george_polimor_2023}.

\textbf{Changelog Monitoring and Event Detection.}
Unlike local file systems that expose OS-level event primitives such as {inotify}~\cite{love_kernel_2005}, {kqueue}~\cite{lemon_kqueue_2001}, and FSEvents \cite{apple_fsevents_2012}, HPC file systems provide dedicated telemetry interfaces. 
FSMonitor~\cite{paul_fsmonitor_2019} unifies event sources across local (Linux, macOS, Windows) and Lustre file systems by defining a common event schema and aggregating notifications from them. 
While FSMonitor can ingest tens of thousands of events per second, its architecture remains centralized: collecting events %via local ZeroMQ
and storing them in a single MySQL database without additional processing or exposing an external, queryable state index. Brindexer’s re-indexer~\cite{paul_efficient_2020} listens to changelogs for modified directories but applies index updates in periodic, batch-oriented runs rather than maintaining a continuous real-time stream.

\section{Summary}
\label{sec:conclusion}

Icicle is a comprehensive HPC file system metadata indexing framework that supports both static snapshot ingestion and real-time event monitoring. 
Our evaluation shows that the snapshot pipeline accommodates heterogeneous metadata sources and scales with resources to process larger snapshots, the quantile sketch maintains consistently low error while significantly improving performance, and the real-time monitor scales linearly with file system metadata parallelism. % (MDTs and number of filesets).
The web interface unifies access to indexed metadata with a graphical query builder and interactive visualizations of storage usage. Icicle supports efficient metadata ingestion, fulfilling the requirements for modern HPC file system management.

\section*{Acknowledgment}

We thank the teams of the Diaspora Project and Globus for their valuable comments and feedback. This work was supported in part by the Diaspora Project, funded by the U.S. Department of Energy, Office of Science, Office of Advanced Scientific Computing Research, under Contract DE-AC02-06CH11357, and by the Globus Search Project, funded by the National Science Foundation under Award 2411188.

\bibliographystyle{IEEEtran}
\bibliography{bib/haochen-refs,bib/refs,bib/websites}
\end{document}